\input{psfig.sty}
\documentclass[11pt,a4paper]{article}
\usepackage{jcappub}

\title{Non--power law behavior of the radial profile of phase--space density of halos}
\author[a]{A. Del Popolo}
\affiliation{Dipartimento di Fisica e Astronomia, University Of Catania, \\
Viale Andrea Doria 6, 95125 Catania, Italy}
\emailAdd{adelpopolo@oact.inaf.it}
\abstract{We study the pseudo phase--space density, $\rho(r)/\sigma^3(r)$, of $\Lambda$CDM dark matter halos with and without baryons (baryons+DM, and pure DM), 
by using the model introduced in Del Popolo (2009), which takes into account the effect of dynamical friction, ordered and random angular momentum, baryons adiabatic contraction and dark matter baryons interplay. We examine the radial dependence of $\rho(r)/\sigma^3(r)$ over 9 orders of magnitude in radius 
for structures on galactic and cluster of galaxies scales. We find that $\rho(r)/\sigma^3(r)$ is approximately a power--law only in the range of halo radius 
resolved by current simulations (down to 0.1\% of the virial radius) while it has a non--power law behavior below the quoted scale, with inner profiles changing with mass. The non--power--law behavior is more evident for halos constituted both of dark matter and baryons while halos constituted just of dark matter and with angular momentum chosen to reproduce a Navarro-Frenk-White (NFW) density profile, are characterized by an approximately power--law behavior. 
The results of the present paper lead to conclude that density profiles of the NFW type are compatible with a power--law behavior of $\rho(r)/\sigma^3(r)$, while those flattening to the halo center, like those found in Del Popolo (2009) or the Einasto profile, or the Burkert profile,  cannot produce radial profile of the pseudo--phase--space density that are power--laws at all radii. 
The results argue against universality of the pseudo phase--space density and as a consequence argue against universality of density profiles constituted 
by dark matter and baryons as also discussed in Del Popolo (2009).}
\begin{document}
\maketitle

\section{Introduction}

Density profiles of dark matter (DM) halos has been intensively studied in the last decade. While the work of Navarro et al. (1996, 1997) concluded that 
spherically--averaged density profile, $\rho(r)$, of equilibrium structure of dark matter (DM) halos, is well approximated by a universal profile known as the Navarro-Frenk-White (NFW) profile,  
more recently, it has been shown that the functional form of the universal profile is better approximated by profiles whose logarithmic slope, $d \ln \rho/ d \ln r \propto r^{\alpha}$, becomes increasingly shallower inwards (Power et al. 2003; Hayashi et al. 2004 and Fukushige et al. 2004; Navarro et al. 2004; Stadel et al. 2009). 
Moreover, there is discussion whether the 
profile is actually universal or not (Moore et al. 1998; Klypin et al. 2001; Navarro et al. 2004; Fukushige et al. 2004; Merrit et al. 2006; Graham et al. 2006; Gao et al. 2008). Regardless of which density profile functional form proves to best describe N-body halos, the underlying physics that drives the halos to have this shape is not yet
fully understood. 

Syer \& White (1998) and Nusser \& Sheth (1999) claimed that the universal profile is a result of hierarchical clustering by mergers of smaller halos into bigger ones. However, Moore et al. (1999) performed N-body simulations with a cut-off in the power spectrum at small scales
and also obtained halos with cuspy density profiles. Similarly Wang \& White (2008) studied the properties of the first generation of haloes in Hot dark Matter dominated universe and compared the properties of the last with CDM ones. They concluded that mergers  and substructure do not play a pivotal role in establishing the universalities.
Huss et al. (1999a,b) found that
simulations of isolated halos collapsing more or less spherically also
result in universal profiles, thus suggesting that hierarchical
merging is not crucial to the outcome.  

A different approach was taken by Taylor \& Navarro (2001) (hereafter TN01), and Hansen (2004). Instead of considering the radial run of the space density of N-body halos, they measured the quantity $\rho(r)/\sigma^3(r)$\footnote{$\sigma(r)$is the one dimensional radial velocity dispersion}, which has the dimensionality of phase--space density. In spherically--symmetric equilibrium halos $\rho(r)/\sigma^3(r)$ is proportional to the coarse-grained phase-space density, a quantity distinct from the fine--grained phase--space density whose conservation is ensured by the collisionless Boltzman equation (e.g., Dehnen 2005). Taylor \& Navarro (2001) identified that the quantity $Q(r)=\rho/\sigma^3(r)$, which has become known as the pseudo phase-space density, behaves as a power law over 2-3 orders of magnitude in radius inside the virial
radius.
Other studies (e.g., Rasia et al. 2004; Ascasibar et al. 2004), have confirmed the scale-free nature of $Q(r)$, and their
results indicate that its slope lies in the range $\alpha=1.90\pm0.05$. This property is remarkable since the density $\rho(r)$ nor the
velocity dispersion $\sigma(r)$ separately show a power-law behavior.
More recently, Ludlow et al. (2011) calculated $Q(r)$
of Einasto halos, which accurately describe the spherically-averaged density profiles of cold dark matter halos. They concluded that $Q(r)$ 
of Einasto halos should be close to power laws over a wide range of radii. However, very close to center, $Q(r)$
for values of $\alpha$ typical of CDM halos deviate significantly from a power law.

%
%
Discussions dealing with the origin of the identified density distributions can be broadly divided into two main theses along basic nature (the profiles result from the 
initial collapse: Lokas \& Hoffman 2001; Nusser 2001; Hiotelis 2002; Kazantzidis et al. 2005; Austin et al. 2005, Barnes et a. 2006) or nurture (they arise from cumulative effect repeated mergers or interactions: Syer \& White 1998; Dekel et al. 2003; Arad et al. 2004; Hoffman et al. 2007; Peirani \& de Freitas-Pacheco 2007) lines.
Despite the insights obtained in the previous studies, the origin for such universality of $Q(r)$is not yet understood, and the question of how this quantity relates to the true coarse-grained phase-space density has been investigated in several papers (Hoffman et al. 2007; Vass et al. 2008; Maciejewski et al. 2009; Sharma \& Steinmetz 2006). 

%
%
Nevertheless, the universality of $Q(r)$ is intriguing since there is no a priori reason to expect that $\rho(r)$ and $\sigma(r)$ should change in such a way as to preserve their ratio over more than $2.5$ orders of magnitude 
in radius, regardless of mass and background cosmology (Taylor \& Navarro 2001, Ascasibar et al. 2004, Rasia et al. 2004, Hoffman et al 2007), and for this reason it has been studied and continues to be studied. 

%
%
Some findings have called into question the universality of $\rho/\sigma(r)^3$. For instance, Schmidt et al. (2008) have shown that simulated halos are better fit by a
different power--law relation, and Ma et al. (2009), found that $\rho/\sigma(r)^3$ is approximately a power law only over the limited range of
halo radius resolvable by current simulations.

We should also add that all the previous quoted analysis 
do not study the possible effects produced by the presence of baryons on $\rho/\sigma(r)^3$, whose effect is to shallow (El-Zant et al. 2001, 2004; Romano-Diaz et al. 2008) and to steepen (Blumenthal et al. 1986; Gnedin et al. 2004; Klypin et al. 2002) the dark matter profile. In real galaxies, the two quoted effects combine with the result of giving rise to density profiles which are different from those predicted in N-body simulations (Del Popolo 2009). In collisionless N-body simulations, this complicated interplay between baryons and dark matter
is not taken into account, because it is very hard to include the effects of baryons in the simulations. However, in order to have a clear view of what simulations can tell about the inner parts of the density profiles and $\rho/\sigma(r)^3$, it necessary to run N-body simulations that repeat the mass modeling including a self-consistent treatment of the baryons and dark matter component. The question is whether or not baryon-DM interactions are universal, or depend on scale - that could be answered by simulations in the next few years.

In the present paper, we focus on the study of the phase-space density proxy $\rho/\sigma^3$ in the case of dark matter and baryons halos, by using the results of Del Popolo (2009).
%
%

The paper is organized as follows: in Section 2, we summarize the Del Popolo (2009) model. In Section 3, we discuss the results. Finally, Section 4 is devoted to conclusions.

\section{Summary of the method}

In the following, we summarize the method used in the present paper, which is fully described in Del Popolo (2009).
In the secondary infall model (SIM) of 
Gunn \& Gott (1972), a bound mass shell of initial comoving radius $x_i$ will expand to a maximum radius $x_m$ (named apapsis or turnaround radius $x_{ta}$). As successive shells expand to their maximum radius, they acquire angular momentum and then contract on orbits determined by the angular momentum. Dissipative physics and the process of violent relaxation will eventually intervene and convert the kinetic energy of collapse into random motions (virialization). 
In SIM, knowing the initial comoving radius $x_i$, the mean fractional density excess inside the shell, as measured at
current epoch $t_0$, assuming linear growth, namely $\overline \delta_i$, and the density parameter $\Omega_i$, it is possible to obtain 
the final time averaged radius of a given Lagrangian shell (Peebles 1980).
If mass is conserved and each shell is kept at its turn-around radius, one can easily obtain the shape of the density profile 
(Peebles 1980; Hoffman \& Shaham 1985 (hereafter HS); White \& Zaritsky 1992).
In reality, after reaching maximum radius, a shell collapses and will start oscillating and it will contribute to the inner 
shells and so even energy is not  an integral of motion anymore. The effect of the in-falling outer
shells on the dynamics of a given shell can be described assuming that the collapse is ``gentle". One can assume that the 
potential well near the center varies adiabatically (Gunn 1977, Filmore \& Goldreich 1984 (FG84)), which means that a shell near the center makes many oscillations before the potential changes significantly (Gunn 1977, FG84, Zaroubi \& Hoffman 1993).
If a shell has an apapsis radius (i.e., apocenter) $x_m$ and initial radius $x_i$, then the mass inside $x_m$ is obtained summing the mass contained in shells 
with apapsis smaller than $x_m$ (permanent component, $m_p$) and the contribution of the outer shells passing
momentarily through the shell $x_m$ (additional mass $m_{add}$),
\begin{equation}
m_T(x_m)=m_p(x_m)+m_{add}(x_m)
\label{eq:mpp}
\end{equation}
The additional mass, $m_{add}$, is proportional to the probability to find the shell with apapsis $x$ inside radius $x_m$, calculated as the ratio of the time the outer shell (with apapsis $x$) spends inside radius $x_m$ to its period. The quoted probability depends on the radial velocity of the shell which can be obtained by integrating the equation of motion of the shell. The final density profile can be obtained in terms of the density profile at turn-around, the collapse factor, and the turn-around radius (Eq. A18, Del Popolo 2009).
In order to calculate the density profile it is necessary to calculate the initial overdensity $\overline \delta_i (x_i)$. 
This can be calculated when the spectrum of perturbations is known. 
As in HS we assume, according to the hierarchical scenario of structure formation, 
that haloes collapse around maxima of the smoothed density field.  
The density profile of a proto-halo is taken to be the profile of a peak in a density field described by the Bardeen et al. (1986) power spectrum, as is
illustrated in Del Popolo (2009), Fig. 6.
The model takes into account angular momentum, dynamical friction, and baryons adiabatic contraction. 
There are two sources of angular momentum of collisionless dark matter: (a) bulk streaming motions, and (b) random tangential motions.
The first one (ordered angular momentum (Ryden \& Gunn 1987 (RG87)) arises due to tidal torques experienced by proto-halos.
The second one (random angular momentum (RG87)) is connected to random velocities (see RG87 and the following of the present paper).
We took into account both types of angular momentum: random $j$, and ordered, $h$. Type (a), 
is got obtaining the rms torque, $\tau (r)$, on a mass shell and then calculating the total specific angular momentum, $h(r,\nu )$, acquired during expansion by integrating the torque over time (Ryden 1988a (hereafter R88), Eq. 35). The random part of angular momentum was assigned to protostructures according to Avila-Reese et al. (1998) scheme, who expressed the specific angular momentum $j$ through the ratio $e_0=\left( \frac{r_{min}}{r_{max}} \right)_0$, where $r_{min}$ and $r_{max}$ are the maximum and minimum penetration of the shell toward the center, respectively, and left this quantity as a free parameter. Processes related to mergers and tidal forces that could produce tangential perturbations in the collapsing matter were implicitly considered in their model trough the free parameter $e_0$. According to them, the detailed description of these processes is largely erased by the virialization process, remaining only through the value of $e_0$, which then they fixed to $e_0=0.3$. 
The value $e_0 \simeq 0.2$ gives density profiles very close to the NFW profile (Avila-Reese et al. 1998, 1999).
We took into account dynamical friction by introducing the dynamical friction force in the equation of motion (Eq. A14 in Del Popolo 2009). 
Dynamical friction was calculated dividing the gravitational field into an average and a random component generated by the clumps constituting hierarchical universes.
The shape of the central density profile is influenced by baryonic collapse: baryons drag dark matter in the so called adiabatic contraction (AC) 
steepening the dark matter density slope. Blumenthal et al. (1986) described an approximate analytical  model to calculate the effects of AC. More recently Gnedin et al. (2004) proposed a simple modification of the Blumenthal model, which describes numerical results more accurately. 
For systems in which angular momentum is exchanged between baryons and dark matter (e.g., through dynamical friction),
Klypin et al. (2002) introduced a modification to Blumenthal's model. 
The adiabatic contraction was taken into account by means of Gnedin et al. (2004) model and Klypin et al. (2002) model taking also account of exchange of angular momentum between baryons and dark matter.

Our method of halo formation has considerable flexibility with direct control over the parameter space of initial conditions differently from 
numerical simulations which yield little physical insight beyond empirical findings precisely because 
they are so rich in dynamical processes, which are hard to disentangle and 
interpret in terms of underlying physics.


In order to calculate $\rho(r)/\sigma^3(r)$, as Williams et al. (2004), we determine for different halos their density profiles, $\rho(r)$ and $\sigma(r)$, by means of Del Popolo (2009).

\section{Result and discussions}

As previously discussed in the introduction, several studies have found that $\rho(r)/\sigma^3(r)$ behaves as a power law over 2-3 orders of magnitude in radius inside the virial radius. This finding is unexpected because the density profile, $\rho(r)$, undergoes a considerable slope change in the same radial interval. The scale-free nature of the phase-space density implies that the double logarithmic slope of the velocity dispersion changes in such a way so as to offset the change in the density profile slope. 
A question that could be asked is the following: does the power--law nature found is a real characteristic of 
all equilibrium N-body halos, at galactic scales or cluster of galaxies scales, or this kind of behavior has been observed since $\rho(r)/\sigma^3(r)$
has been studied in a limited range of halo radius and without taking into account the effect of baryons?  
In order to answer this question and seeing if the same behavior is valid for halos of different masses, galaxies or clusters of galaxies, we shall use the SIM model summarized in the previous section and fully described in Del Popolo (2009) to calculate $\rho(r)/\sigma^3(r)$ in a wider radial range than that resolved by current simulations. Using Del Popolo (2009), we generate two sets of halos with galaxy and cluster scale masses. Within each of the two sets, we study three different cases: in the case A we take into account all the effects included in Del Popolo (2009), namely angular momentum, dynamical friction, baryons, adiabatic contraction of dark matter; in the case B there are no baryons and dynamical friction;
in case C only angular momentum is taken into account reduced in magnitude as in Del Popolo (2009) in order to reproduce the angular momentum of N-body simulations (dashed line in Fig. 2 of Del Popolo 2009) and a NFW profile (solid histogram in Fig. 2 of Del Popolo 2009). 

For what concerns the case C, we recall that in Del Popolo (2009), we performed an experiment similar to that performed by Williams et al. (2004) namely we reduced the magnitude of the 
angular momentum, of a factor of 2, and that of the dynamical friction force, 
of a factor 2.5 with respect to the typical values calculated and used in the model in order to reproduce angular momentum of N-body simulations and the NFW profile. 

\subsection{Pseudo--phase--space density for galaxies and clusters}

In Fig. 1a, we plot $\rho(r)/\sigma^3(r)$ with respect to radius, over 9 orders of magnitude, for a galaxy of $10^9 M_{\odot}$. The solid, dotted and long--dashed lines represent, respectively: (a) the slope in the case A; b) the slope in the case B; (c) the slope in the case C. The short--dashed line represents the slope $-1.9$ found in several studies.\footnote{As reported in the introduction some studies, e.g. Ascasibar et al. 2004, showed that the slope $\alpha= 1.90 \pm 0.05$, compatible with Taylor \& Navarro (2001) value of $\alpha=-1.875$} 
Fig. 1b shows a zoom--in view of the portion of the left figure that is resolvable by current simulations: 
$0.003 \leq /r_{-2} \leq 30$\footnote{$r_{-2}$ is the radius at which the logarithmic slope of the
space density is -2}. Fig. 1 illustrates that $\rho(r)/\sigma^3(r)$ is not a power--law in $r$ both for case A and B. In the case A and B halos deviate strongly from a power--law at small radius. The zoom--in panels shows, however that the slope of $\rho(r)/\sigma^3(r)$ happen to be quite close to $-1.9$  over the limited range of $r/r_{-2} \simeq 0.001$ to 10 that is resolvable by current simulations. The deviations only starting to show up at the smallest radius $r/r_{-2} \simeq 0.001$ near the simulation resolution limit. It is therefore not surprising that the power-law behavior of $\rho(r)/\sigma^3(r)$ continues to appear to be valid even though the latest simulations find the Einasto form to be a better fit for $\rho(r)$ than GNFW (generalized NFW).
If for $\geq 0.001$ the $\rho(r)/\sigma^3(r)$ is more or less a power--law, the important point to note, is that at radius smaller than $r/r_{-2} \leq 0.001$, 
($10^{-7} \leq r/r_{-2} \leq 0.001$), $\rho(r)/\sigma^3(r)$ deviates far away from a pure power law.
For our case A and B, 
the shape of $\rho(r)/\sigma^3(r)$ flattens continuously towards the halo center.
This is not unexpected since the corresponding 
density profile, as shown in Del Popolo (2009), flattens going to the inner part of the halo, where has a flat profile. 
As already reported, the density profile corresponding to the case C are characterized by the fact that we take into account only angular momentum reduced as in Del Popolo (2009) in order to reproduce a NFW profile with inner slope $\rho \propto r^{-1}$.  
The corresponding $\rho(r)/\sigma^3(r)$ profile (long--dashed line in Fig. 1a) has an approximatively power--law behavior with slope $\rho(r)/\sigma^3(r) \propto r^{-1.9}$ 
in agreement with several of the results in literature (e.g., TN01).  
It is worth noting, however, that even the NFW halo
shows wiggles in the corresponding $\rho(r)/\sigma^3(r)$ profile; that is, NFW haloes
have not an exact power-law $\rho(r)/\sigma^3(r)$.  
%

At large values of radius, $r/r_{-2} \geq 1$ there is also a deviation of $\rho(r)/\sigma^3(r)$ from a power--law. The reason of this deviation can be explained by means of Jeans equation, as done in Williams et al. (2004), and it is due to the fact that at large radii the equilibrium condition is not well satisfied. This apparently leads to a break in the scale--free behavior of $\rho(r)/\sigma^3(r)$, at large radii.

In Fig. 2, we plot $\rho(r)/\sigma^3(r)$ with respect to radius, in the case of a cluster of galaxies of $10^{14} M_{\odot}$. Here, the solid line, the dotted line, and the dashed line represents case A, case B, and case C. The behavior of the phase--space proxy is similar to the case of the galaxy. Also in this case, at small and large radii $\rho(r)/\sigma^3(r)$ is not a power--law, but in a different radius range with respect to galaxies, namely $10^{-7} \leq r/r_{-2} \leq 0.01$ and $r/r_{-2} \leq 10$. 
Moreover, the $\rho(r)/\sigma^3(r)$ slope of both case A and B, for small radii, are steeper than in the case of the galaxy, and the slope of curve relative to the NFW profile (dashed line) is slightly less steep, namely $\rho/\sigma^3 \propto r^{-0.8}$, in agreement with Williams et al. (2004) results. 

In order to explain why the inner $\rho(r)/\sigma^3(r)$ profiles are not power--laws and why the slopes are steeper in the cluster of galaxies case, we have to recall how the density profiles, $\rho(r)$, are formed in Del Popolo (2009).
The differences in slopes with mass for the three plotted cases (A, B, and C) can be explained as follows. In case A, baryons, dynamical friction and angular momentum are present. The final density profile and final slope is determined by the interplay of these three factors. Let's see how each of these factors act in shaping the profile and how they interplay.
For what concerns angular momentum, we have to note that less massive objects are generated by peaks of smaller height, which acquire more angular momentum.
The angular momentum sets the shape of the density profile at the inner regions. For pure radial orbits, the core is dominated
by particles from the outer shells. As the angular momentum increases, these particles remains closer to the maximum radius,
resulting in a shallower density profile. Particles with smaller angular momentum will be able to enter the core but with a reduced
radial velocity compared with the purely radial SIM. For some particles, the angular momentum is so large that they will never
become unbound. Summarizing, particles of larger angular momenta are prevented from coming close to the halo's center and
so contributing to the central density, which has the effect of flattening the density profile. 
%
%
The effects of dynamical friction can be interpreted in two different fashions: (a) an increase in the dynamical friction force
is very similar to changing the magnitude of angular momentum (see Fig. 11 of Del Popolo 2009) with the final result of producing shallower profiles; (b) dynamical friction can act on gas moving in the background of dark matter particles, dissipating the clumps orbital energy and depositing it in the dark matter with the final effect of giving rise to a flatter profile
(similarly to El-Zant et al. (2001); El-Zant et al. 2004; Tonini, Lapi \& Salucci 2006 (TLS); Romano-Diaz et al. 2008). 
In order to make more clear the previous discussion, in Fig. 3a, we show the effect of changing angular momentum and dynamical friction (we plot the joint effect of angular momentum and dynamical friction because they produce a similar effect on the density profile) on a profile of $10^{10} M_{\odot}$ (solid line), obtained as in Del Popolo (2009). The dashed (dotted) line represents the density profile when increasing (decreasing) angular momentum and dynamical friction of a factor of 1.5. As shown, increasing the magnitude of angular momentum and dynamical friction produces a flattening of the profile. Decreasing them leads to the opposite effect.
Baryons have another effect, at an early redshift, the dark matter density experiences the adiabatic contraction by baryons producing a slightly more cuspy profile. This last is overcome from the previous two effects. As shown by Fig. 11 of Del Popolo (2009), the magnitude of dynamical friction effect is a bit larger than that due to angular momentum and that these two effects add to improve the flattening of the profile.
In Fig. 3b, we plot the effect of adiabatic contraction on the density profiles plotted in Fig. 3a. As shown dashed and dotted lines show that adiabatic contraction, as expected, produces a steepening of the profile. 

The quoted effects act in a complicated interplay. Initially, at high redshift (e.g., $z=50$), the density profile is in the linear regime. The profile evolves to
the non-linear regime, and virialize.
At an early redshift, (e.g. $z \simeq 5$, for dwarf galaxies), the dark matter density experiences the adiabatic contraction by baryons producing a slightly more cuspy profile.
The evolution after virialization is produced by secondary infall, two-body relaxation, dynamical friction and angular momentum. 
Angular momentum, as described, contributes to reduce the inner slope of density profiles by preventing particles from reaching halo's center, while dynamical friction 
dissipate the clumps orbital energy and deposit it in the dark matter with the final effect of erasing the cusp (similarly to El-Zant et al. (2001); El-Zant et al. 2004; TLS; Romano-Diaz et al. 2008).
The cusp is slowly eliminated and within $ \simeq 1$ kpc a core forms, for objects of the mass of dwarf-galaxies. 
It is now clear why going from a model which takes into account baryons, dynamical friction, and angular momentum (solid line, case A) to one taking account just of angular momentum (dashed line, case B) and to one taking account just of angular momentum reduced to reobtain N-body simulations angular momentum,
one obtains so different behavior of inner density profile slopes.

\begin{figure}[tbp]
\centerline{\hbox{(a)
\hspace{2cm} \psfig{figure=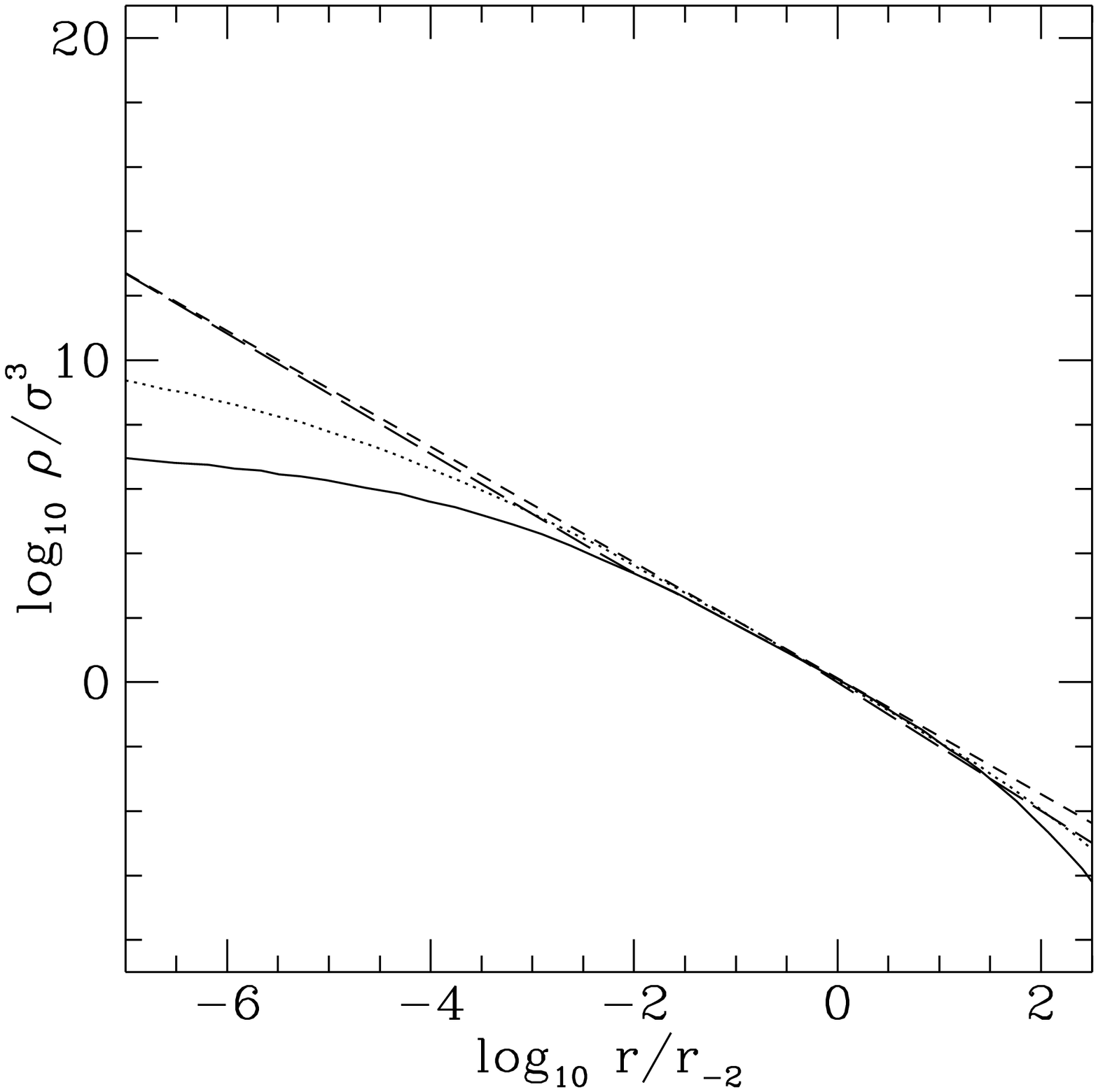,width=8cm} (b)
\hspace{0cm} \psfig{figure=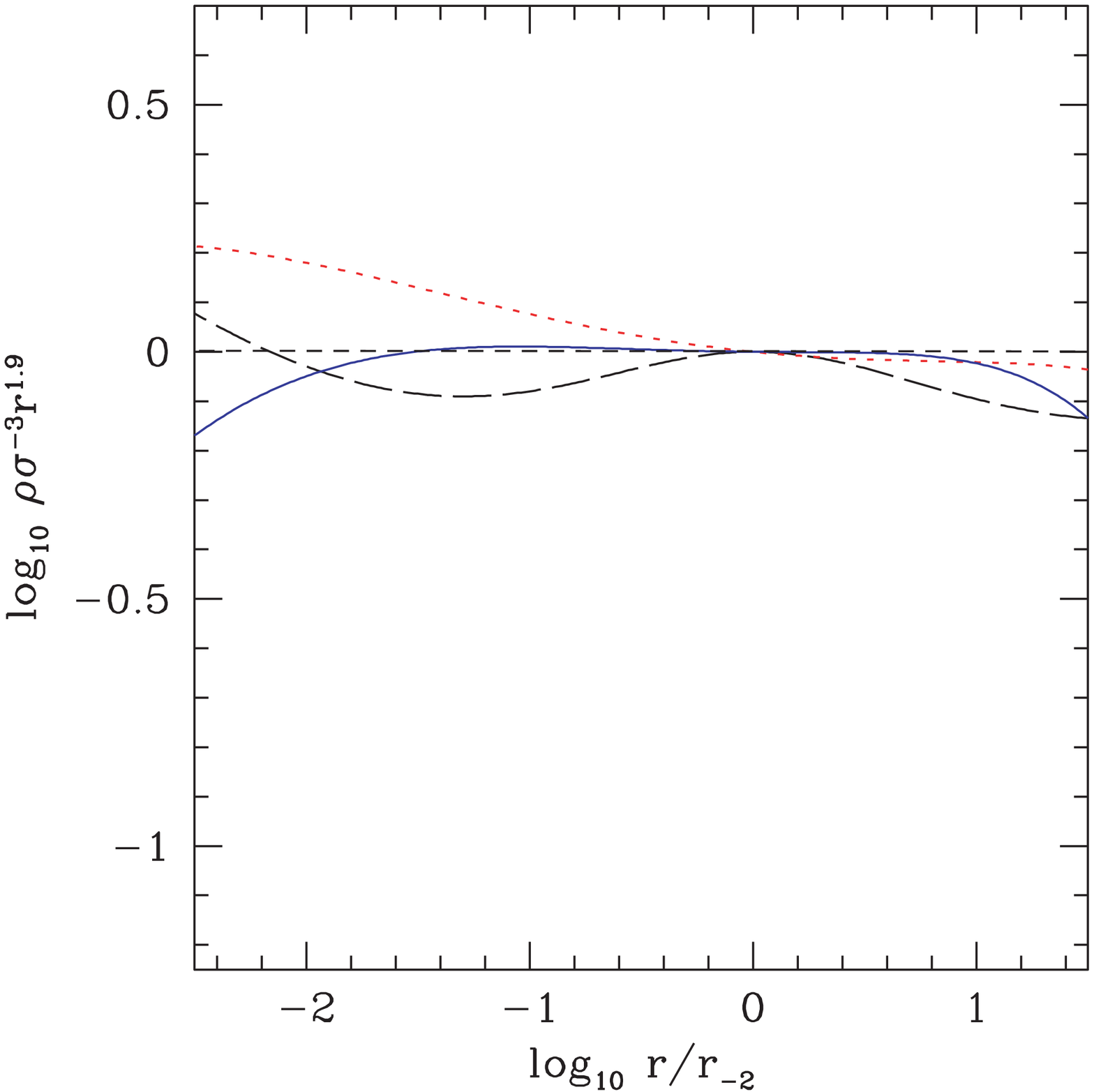,width=8cm}  
}}
\caption{Radial profiles of the pseudo--phase--space density for a $10^9 M_{\odot}$ galaxy. Panel (a): solid line, dotted line, and long--dashed line represent, respectively, the pseudo--phase--space density for the case A, B, and C described in the paper. The short--dashed line represents the slope $-1.9$ found in several studies (see the introduction). Panel (b) show a zoom--in view of the portion of the left figure, multiplied by $r^{1.875}$, that is resolvable by current simulations. Symbols are as in panel (a).
}
\end{figure}

\begin{figure}[tbp]
\centerline{\hbox{
\hspace{2cm} \psfig{figure=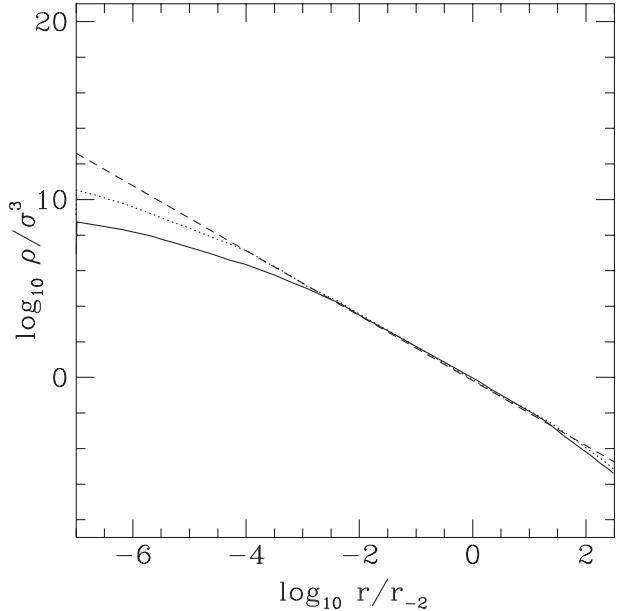,width=8cm} 
}}
\caption{Radial profiles of the pseudo--phase--space density for a $10^{14} M_{\odot}$ cluster. Solid line, dotted line, and long--dashed line represent, respectively, the pseudo--phase--space density for the case A, B, and C described in the paper. 
}
\end{figure}

The main reason of the difference of behavior between clusters and galaxies is due to the fact that in the case of clusters the virialization process starts much later with respect to galaxies. In the case of galaxies, the profile strongly evolve after virialization through the processes previously described. In the case of dwarf galaxies of $10^9 M _{\odot}$, we showed in Del Popolo (2009) that the profile virializes at $z \simeq 10$ and from this redshit to $z=0$ its shape continues to evolve, except at $z \simeq 5$ when adiabatic contraction steepen the profile. In the case of a cluster of $10^{14} M _{\odot}$, 
the profile virializes at $z \simeq 0$ (Del Popolo 2009) and, as a consequence, the further evolution observed in galaxies cannot be observed in clusters. 
\begin{figure}
\hspace{-1.5cm} 
(a)
\includegraphics[width=84mm]{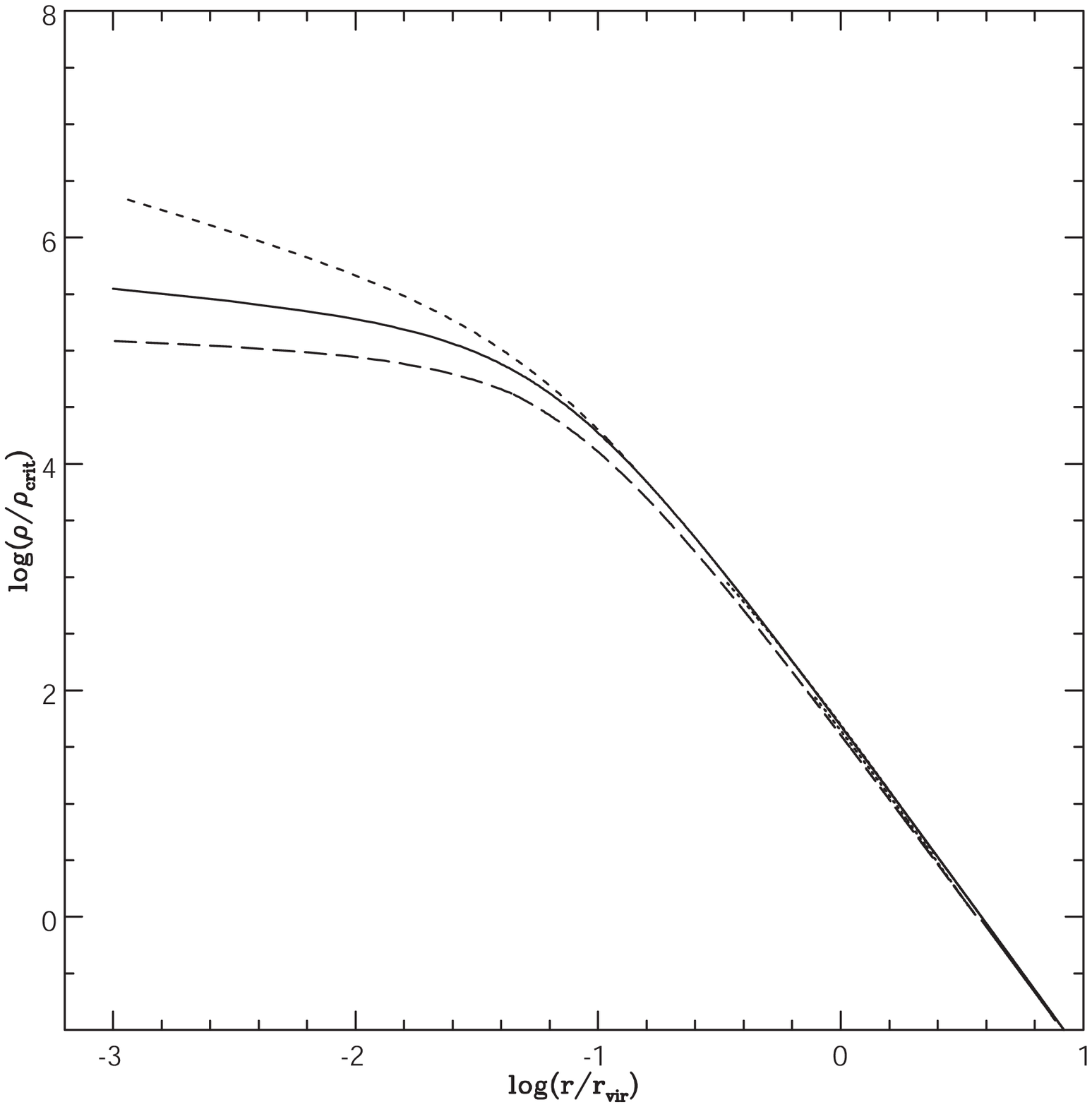} (b) 
\includegraphics[width=84mm]{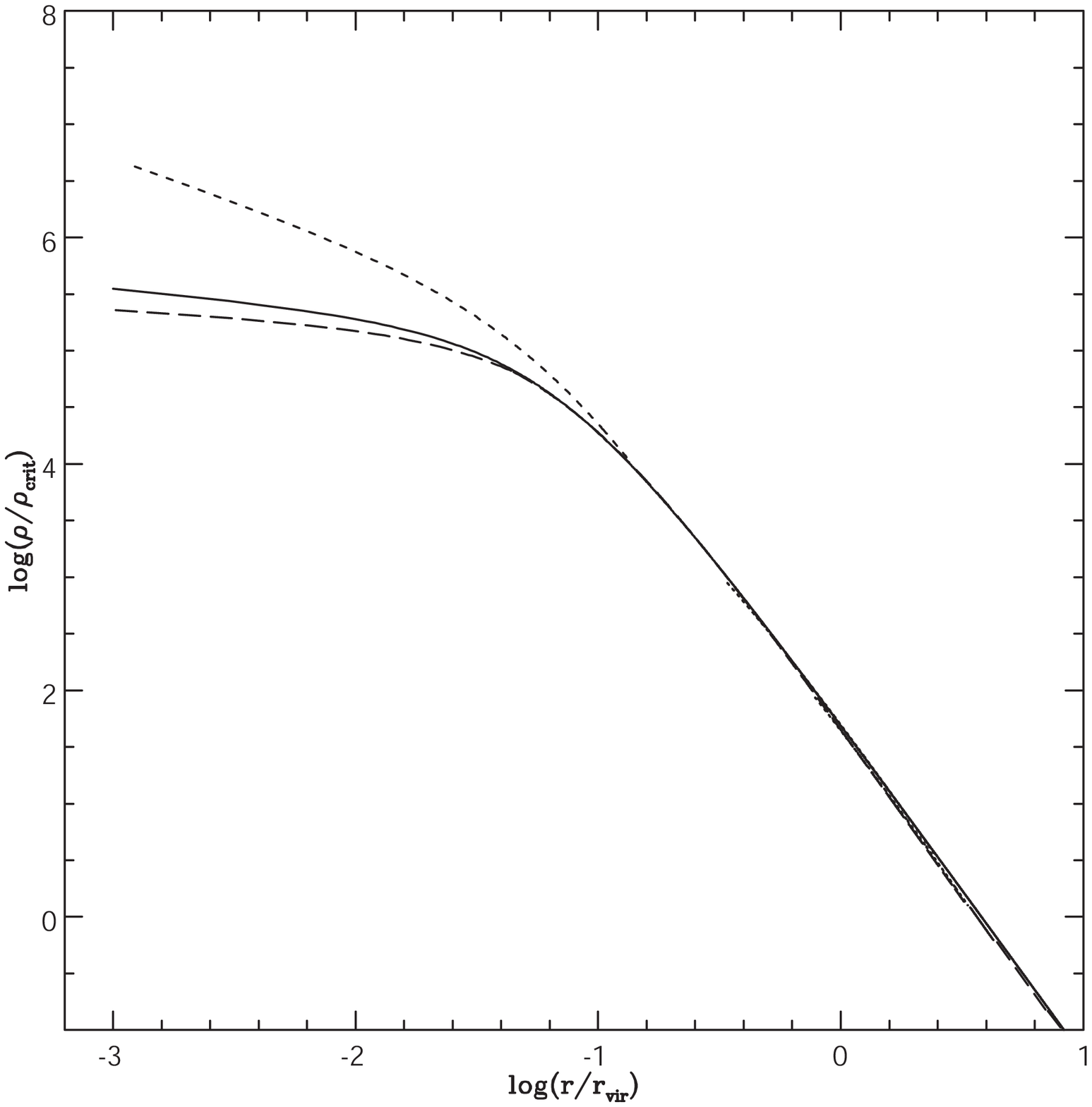} 
\caption{Panel (a): Change of density profile of DM halos with angular momentum. The solid line represents the density profile of an halo of $10^{10} M_{\odot}$ calculated as in Del Popolo (2009). Dashed (dotted) line represents the effect of increasing (decreasing), of a factor of 1.5, the value of angular momentum and dynamical friction, that gave rise to the profile represented by the solid line. 
Panel (b): Change of density profile of DM halos of panel (a) due to adiabatic contraction.
}
\end{figure} 
Summarizing, Fig. 1 and Fig. 2 show that $\rho(r)/\sigma^3(r)$ is not a power--law as shown in some previous studies reported in the introduction. Our results are in line with some recent work that has called into question the universality of $\rho/\sigma(r)^3$. For instance, Schmidt et al. (2008) has advocated that individual simulated haloes are better fit by a generalized power-law relation that is not necessarily $\rho/\sigma(r)^3$:
\begin{equation}
\frac {\rho}{\sigma_D^{\epsilon}} \propto r^{-\alpha},
\label{rsignew}
\end{equation}
where $\sigma_D = \sigma_r\sqrt{1 + D\beta}$, and $D$ parameterizes a
generalized $\sigma_D$; for instance, $D=0$ 
and $-1$ correspond to $\sigma_D=\sigma(r)$, and $\sigma_t$, respectively\footnote{Note that $\sigma(r)$, and $\sigma_t$, are the one dimensional 
radial and tangential velocity dispersions.}.
Schmidt et al. (2008) showed that the best-fit values of $(D, \epsilon, \alpha)$
differ from halo to halo, and as a set, they roughly follow the linear relations $\epsilon=0.97 D + 3.15$ and $\alpha=0.19 D + 1.94$.  For $\sigma=\sigma(r)$ (i.e. $D=0$), the optimal relation is $\rho/\sigma(r)^{3.15} \propto r^{-1.94}$, which is consistent with the reported behavior of $\rho/\sigma(r)^3$ in $N$-body simulations within error bars.
Similarly, Ma et al. (2009), examining the radial dependence of $\rho/\sigma(r)^3$ over 12 orders of magnitude in radius by solving the Jeans equation for a broad range of input $\rho$ and velocity anisotropy $\beta$, found that $\rho/\sigma(r)^3$ is approximately a power law only over the limited range of
halo radius resolvable by current simulations (down to $\sim 0.1$\% of the virial radius), and $\rho/\sigma(r)^3$ deviates significantly from a power-law
below this scale for both the Einasto and NFW density profiles, $\rho(r)$.  

\subsection{Universality of the phase-space density profile}

Another issue to discuss is the interrelation between the universality of the pseudo--phase--space density and that of the halos density profiles. As previously discussed, there are different methods to analyze dark matter halos structure. A standard approach involves investigating the halos density profiles. 
%
%
Few theoretical attempts have been made to understand the origin of this density profile (e.g., Salvador-Sol\'e et al. 2007; Henriksen et al. 2007), with varying level of success. 
As we already know, a completely different approach is to search for simple phenomenological relations in the numerical simulations, such as finding straight lines in some parameter space as done for the first time by Taylor \& Navarro (2001), and Hansen (2004). 
The scale-free nature of $\rho(r)/\sigma^3(r)$ represents a novel way of looking at the properties of halos. If this property is ``universal'', it amounts to a hitherto unrecognized constraint on the shape of the density profiles (Austin et al. 2005). Our result confirm this point of view: $\rho(r)/\sigma^3(r)$ profiles flattening towards the halo center are generated by similar density profiles, which have logarithmic slope $\alpha \simeq 0$ for $10^8-10^{9} M_{\odot}$ and $\alpha \simeq 0.6$ for $M \simeq 10^{14} M_{\odot}$ (Del Popolo 2009). At the same time, our main result is that $\rho(r)/\sigma^3(r)$ is not universal if studied in the appropriate radius range, and similarly we expect that halos density profiles are not universal, because their inner part should depend on mass, as the $\rho(r)/\sigma^3(r)$ profiles, and should also flatten towards the halo center, showing flat cores in the center of the halo (as shown in Del Popolo 2009). 
This result is in agreement with several previous ones, described in the reminder.

Another author who addressed the problem of the universality of the phase-space
density profile, is Knollmann et al. (2008), finding a mixed answer. On the one hand it
has been very robustly shown that indeed a power law profile provides a
good fit in all models considered in their study.  Yet, the power law slope
$\alpha$ has been found to vary with power spectrum index $n$.  Given
our lack of understanding of the origin of the phase-space density power
law it is interesting to trace the reason for the dependence of the
slopes $\alpha$ on $n$. According to Knollmann et al. (2008), at least part of
that dependence is attributed to the concentration parameter variation
with $n$.  However, it is unclear whether there is a direct dependence
of the slope $\alpha$ on $n$.  

The result of the non--universality of $\rho(r)/\sigma^3(r)$ is in agreement with other studies 
on the universality of density profile, $\rho(r)$.
Ricotti's (2003) N-body simulations suggested that the density profile of DM haloes 
is not universal (in agreement with the other quoted studies), but that there are instead shallower cores in dwarf galaxies and steeper cores in clusters.
This leads to the conclusion that density profiles do not have a universal shape. Moreover in agreement with Subramanian (2000), the halo shape at a given mass or spatial scale depends on the slope of the power spectrum at that scale. 
Cen et al. (2004) confirmed Ricotti's result; in addition they identified a redshift dependence of the typical halo profile. 
Graham et al. (2006) and Merrit et al. (2005, 2006) also found a correlation between halo mass and the shape of  the density profile, parameterizing it in terms of the 
S\'ersic profile index.  
There are also observational evidences of a mass dependence of the dark matter density profile.
Simon (2003a,b, 2004), removed the contribution of the stellar disk to the rotation curve of 5 galaxies in  order to reveal the rotation curve of their dark matter halo. They found that the galaxies, NGC2976, NGC6689, NGC5949, NGC4605, NGC5963 have very different values of the slope: $\alpha \simeq 0.01$, 0.80, 0.88, 0.88, 1.28, respectively.
Moreover, observed slopes on galactic scales have large scatter compared to simulations and mean slope shallower than simulations.
In the case of clusters, for both lensing and X-ray studies most authors focus on only one or a few clusters, which of course makes it more difficult to assess the universality of the profiles on an observational foundation.
However, Host \& Hansen (2009) take a sample of 11 highly relaxed clusters and use the measurements of the X-ray emitting gas to infer model-independent mass profiles. They then compare with a number of different models that have been applied as mass profiles in the literature, concluding that there is a strong indication that this inner slope needs to be determined for each cluster individually.
This implies that X-ray observations do not support the idea of a universal inner slope, but perhaps show a hint of a dependence with redshift or mass. 

Before concluding, we want to add a last comment. As already shown, in the limit range of actual N-body simulations $\rho(r)/\sigma^3(r)$ behaves approximately as a power law, even in our study, and this is valid for structures of different mass. This universal behavior (in the cited radii range) has been interpreted in different ways, in different studies, as discussed in the introduction. In this respect, our study is similar to that of Austin et al. (2005), who using semi-analytical extended secondary infall models to follow the evolution of collisionless spherical shells of matter, and by examining the processes in common between 
numerical N-body and semi-analytic approaches, showed  that the power-law behavior of the final phase-space density profile (in the limit range of N--body simulations) cannot be the result of hierarchical merging, but rather it is a robust feature of virialized halos, which have reached equilibrium via violent relaxation.


\subsection{Comparison of our DM density profiles and $\rho(r)/\sigma^3(r)$ 
with simulations}

In order to show the goodness of the model used, we compared in Fig. 4, the results of the same with density profiles  
obtained with dissipationless (Fig. 4a) and SPH (Fig. 4b) simulations. Fig. 4a plots the result of Stadel et al. (2009) dissipationless simulations (solid line), and that of our model (dashed line). In the case of Fig. 4a, the density profile was calculated with the model of the present paper
without taking into account baryons.
Fig. 4b plots the result of our model (dot-dashed line) and galaxy DG1 (solid blue line) and galaxy DG2 (solid black line) 
of Governato et al. (2010) SPH simulations. Fig. 4a and Fig. 4b were obtained by using the same cosmological parameters used by Stadel et al. (2009), and Governato et al. (2010). Fig. 5 compares the velocity anisotropy, $\beta(r)$, obtained with the model of the present paper with Ludlow et al. (2011) results (their Fig. 3). 
In both Fig. 5a and 5b the solid line is $\beta(r)$ obtained by means of the model of the present paper, without taking into account baryons, 
the dot-dashed curves show $\beta(r)$ from equation (5) of Ludlow et al. (2011) 
and the dotted line $\beta(r)$ from Hansen \& Moore (2006). The median anisotropy profile and one-sigma dispersion for each halo subsample, drawn from the Millennium-II simulation, is represented by solid black lines with error bars. 
Fig. 5a refers to the best-fit Einasto parameter $\alpha=0.132$ while Fig. 5b to $\alpha=0.178$.
We also compared (but did not plot)
velocity dispersion with Ascasibar \& Gottl\"ober (2008) finding again a good agreement with their simulations.

In order to show how our results concerning $\rho(r)/\sigma^3(r)$ are in agreement with high-resolution simulations, in the radius range that they studied, 
we compare the results for $\rho(r)/\sigma^3(r)$ 
with recent results of Ludlow et al. (2011), who calculated the pseudo phase--space density for the Einasto profile.

Fig. 6 of the present paper, plots the top panel of Fig. 2 of Ludlow et al. (2011). In Fig. 6, the mean profiles and one-sigma scatter of $Q$ calculated by Ludlow et al. (2011) are shown as solid red lines with error bars. The dotted curve shows Bertschinger $r^{-1.875}$ result, while the solid black line, almost indistinguishable from the dotted line, was calculated with the model of the present paper
without taking into account baryons, in order to be able to compare the result with dissipationless simulations results for $Q$, like those of Ludlow et al. (2011).
Fig. 6 shows a very good agreement of the result of the present paper with those of Ludlow et al. (2011), in their studied radius range $\simeq 10^{-2}$-$10^{1.5}$.\\
\begin{figure}[tbp]
\centerline{\hbox{
\hspace{0cm} \psfig{figure=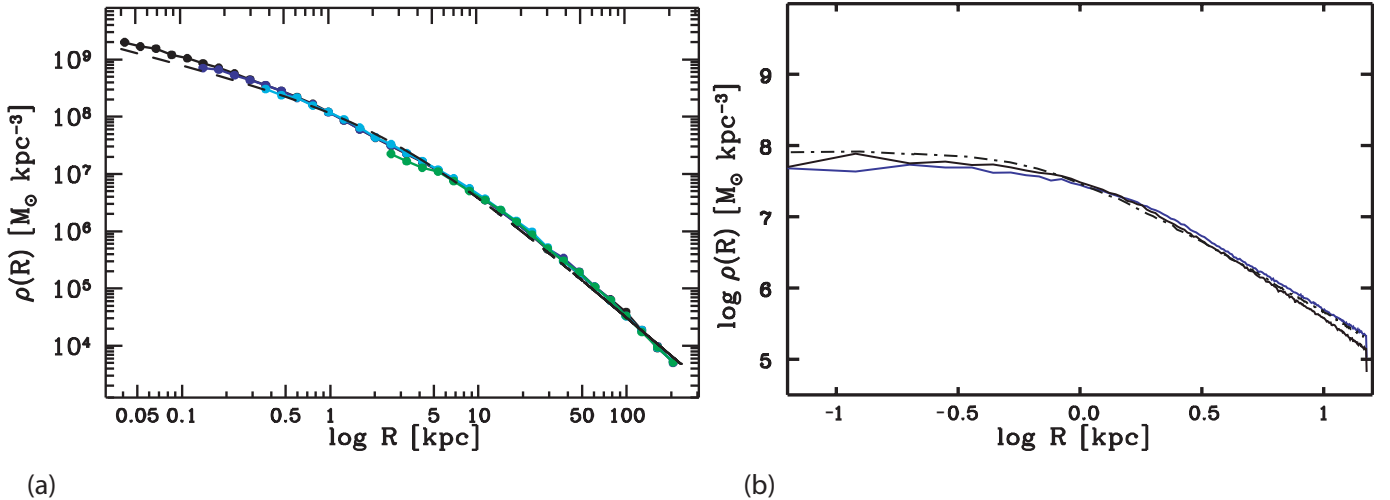,width=18cm} 
}}
\caption{Comparison of the dark matter density profile of our model with simulations. Panel (a) plots the result of Stadel et al. (2009) dissipationless simulations (solid line), and that of our model (dashed line). Panel (b) plots the result of our model (dot-dashed line) and galaxy DG1 (solid blue line) and galaxy DG2 (solid black line) 
of Governato et al. (2010) SPH simulations
}
\end{figure}
Note that $\chi$ in Fig. 6 is the exponent in 
\begin{equation}
Q(r)=\frac{\rho}{\sigma^3}=\frac{\rho_o}{\sigma_o^3}(\frac{r}{r_o})^{-\chi}
\end{equation}
which as previously told was originally reported by Taylor \& Navarro (2001).
\begin{figure}[tbp]
\centerline{\hbox{
\hspace{0cm} \psfig{figure=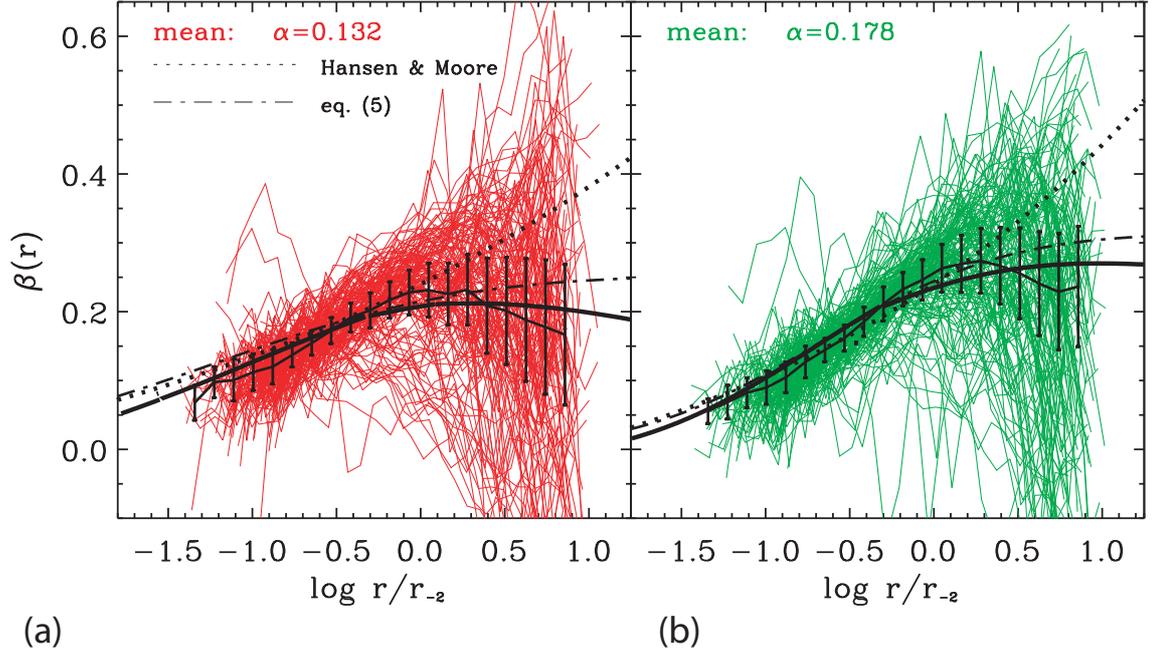,width=15cm}   
}}
\caption{Velocity anisotropy profiles. In this plot, we compare the results of Ludlow et al. (2011) for $\beta(r)$ with our model results.
Both in panel (a) and (b), the solid line is the result of the present paper without taking account of baryons,
the dotted curves show $\beta(r)$  from Hansen \& Moore (2006), while the dot-dashed curve show $\beta(r)$ from equation (5) of Ludlow et al. (2011) assuming the average Einasto density profile. 
The median anisotropy profile and one-sigma dispersion for each halo subsample is represented by solid black lines with error bars. 
Panel (a) refers to the best-fit Einasto parameter $\alpha=0.132$ while panel (b) to $\alpha=0.178$
}
\end{figure}

Fig. 7 shows (in blue) the 
density profiles for three different values of $\chi$, and compares them to Einasto profiles. 
The values of $\alpha$ of the three Einasto profiles shown have been chosen, by Ludlow et al. (2011), to match as closely as possible
the profiles corresponding to the pseudo phase--space density models. Fig. 7 shows (in red) the pseudo phase--space density profiles
of Einasto halos, for three different values of $\alpha$. For $\alpha = 0.1$ and $\alpha = 0.17$ the
corresponding pseudo phase--space density profiles are very well approximated by power laws over the whole plotted radial range. 
\begin{figure}
\centerline{\hbox{
\hspace{0cm} \psfig{figure=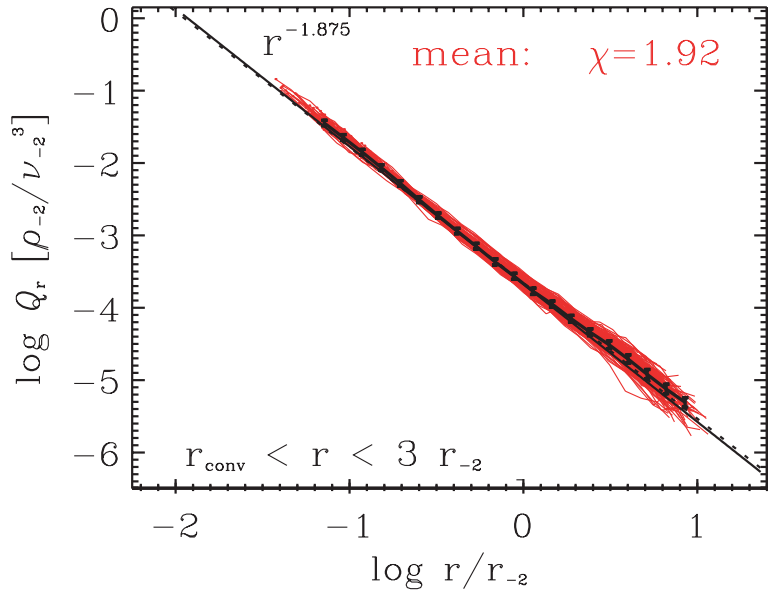,width=10cm}   
}}
\caption[]{Comparison of pseudo-phase-space density profiles of Fig. 2 of Ludlow et al. (2011), with the result of the present paper (solid line). 
}
\end{figure}
Only for larger values of $\alpha$ such as 0.3, clear deviations from a power law are noticeable. 
The thick solid curve in black shows the profile corresponding to the billion-particle Aq-A-1 halo, namely Navarro et. (2010) highest resolution halo.
The solid (dot-dashed) green line marked DP in Fig. 7 plots the pseudo phase--space density profile calculated with the model of the present paper for case B (C).
%
The green dashed line in Fig. 7 shows that the flattening that we obtain at $r/r_{-2} = 10^{-3.7}$ (smallest value plotted by Ludlow et al. 2011 in their Fig. 5b)  
is 
close to the case $\alpha = 0.17$, (typical of $\Lambda$CDM models) plotted in Fig. 5b of Ludlow et al. (2011). 
The flattening in Fig. 5b of Ludlow et al. (2011) for $\alpha = 0.10$ at $r/r_{-2} = 10^{-3.7}$ is almost identical to that of the present paper for case C (green dot-dashed line). 


Fig. 7, shows a very good agreement of the result of my paper with those of Ludlow et al. (2011), in their studied radius range $\simeq 10^{-3.7}$-$10^{1.5}$ both for an Einasto profile with $\alpha=0.17$ and $\alpha=0.10$ . Ludlow et al. (2011) did not probe pseudo-phase-space density profile at a smaller radius, as done in the present paper. At a radius of $\simeq 10^{-3.7}$, minimum radius plotted by Ludlow et al. (2011), a small discrepancy from pure power-law behavior of the pseudo-phase-space density profile starts to be seen, and if one goes to a smaller radius (e.g., $10^{-9}$ as in the submitted paper) the flattening should be larger. 
Moreover, as reported in the Conclusion section of Ludlow et al. (2011): ``significant differences are only expected at radii well inside 1\% of the scale radius, $r_{-2}$, and are therefore beyond the reach of current simulations."

\begin{figure}
\includegraphics[width=90mm]{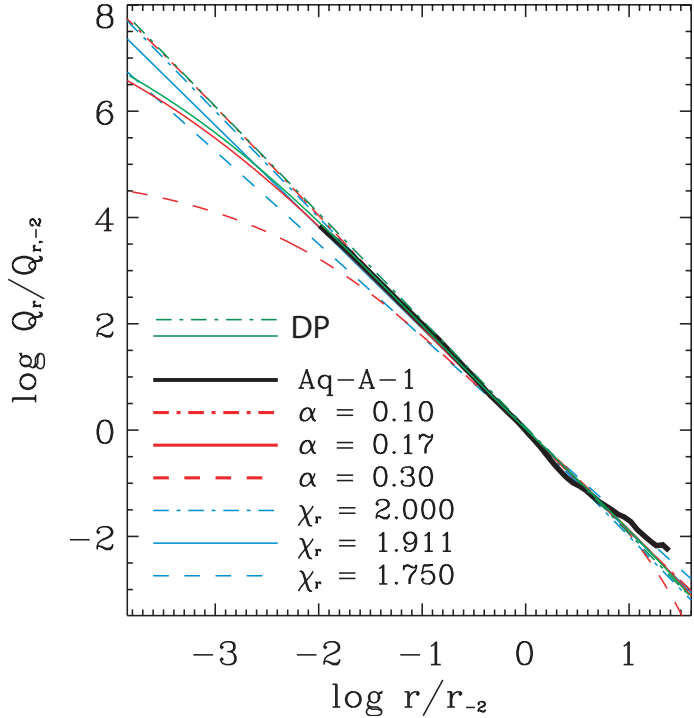}
\caption[]{Comparison of pseudo-phase-space density profiles of Fig. 5b of Ludlow et al. (2011) with those obtained in the present paper. The result of the present paper (solid green line) for case B is in good agreement with pseudo-phase-space density profile of Einasto profile with $\alpha=0.17$ typical of $\Lambda$CDM models and the case $\alpha=0.10$ is almost identical to that of the present paper for case C (green dashed line). 
DP stands for Del Popolo (i.e., the result of the present paper)}
\end{figure}

\section{Conclusions}

Several studies have shown that in spherically--symmetric equilibrium halos, the pseudo phase--space density, $\rho(r)/\sigma^3(r)$, behaves as a power law over 2-3 orders of magnitude in radius inside the virial radius. In order to check this behavior of the quoted quantity, we used the model described in Del Popolo (2009). This last is a SIM 
which takes into account the effect of dynamical friction, ordered and random angular momentum, baryons adiabatic contraction and dark matter baryons interplay. 
We find that $\rho(r)/\sigma^3(r)$ is not in general a power--law for the case A (dark matter and baryons) and B (no baryons) described in the paper. In the radial range probed by current N-body simulations (down to  $10^{-3}$ virial radii), $\rho(r)/\sigma^3(r)$ approximately behaves like a power--law, while for radial scales below the resolution of current simulations, there are significant deviations from a power--law profile. A similar, non power--law behavior is observed at large radii ($> 10$ virial radii). In the paper, we also set the angular momentum and dynamical friction so that the density profile is approximately a NFW profile (case C). In this case $\rho(r)/\sigma^3(r)$ is approximately a power law. The pseudo phase--space density was calculated for structures on galactic, and cluster of galaxy mass scale. The behavior of $\rho(r)/\sigma^3(r)$ observed was similar, but in the case of clusters the slope was steeper in both case A and B. This difference is connected to the different redshift at which the two class of objects formed, larger for galaxies, smaller for clusters, which implies a longer time at disposal of galaxies to evolve. The results of the quoted study are in agreement with those of Schmidt et al. (2008) and Ma et al. (2009). 
We conclude that radial profiles of the pseudo phase--space density corresponding to density profiles which flatten going towards the halo center cannot be power--laws, and the prediction of N-body simulations of a power--law behavior in $\rho(r)/\sigma^3(r)$ is due just to the fact that the pseudo phase--space density is observed only down to a resolution limit of $10^{-3}$ virial radii. 

The results argue against universality of the pseudo phase--space density and as a consequence argue against universality of density profiles constituted 
by dark matter and baryons as also discussed in Del Popolo (2009).




%

\end{document}